# Preservation of Indigenous Culture among Indigenous Migrants through Social Media: the Igorot Peoples

Khavee Agustus Botangen
School of Engineering, Computer, and Mathematical Sciences
Auckland University of Technology, Auckland, New Zealand
khavee.botangen@aut.ac.nz

Shahper Vodanovich
School of Engineering, Computer, and Mathematical Sciences
Auckland University of Technology, Auckland, New Zealand
shahper.vodanovich@aut.ac.nz

Jian Yu
School of Engineering, Computer, and Mathematical Sciences
Auckland University of Technology, Auckland, New Zealand
jian.yu@aut.ac.nz

**Abstract**

*The value and relevance of indigenous knowledge towards sustainability of human societies drives for its preservation. This work explored the use of Facebook groups to promote indigenous knowledge among Igorot peoples in the diaspora. The virtual communities help intensify the connection of Igorot migrants to their traditional culture despite the challenges of assimilation to a different society. A survey of posts on 20 Facebook groups identified and classified the indigenous cultural elements conveyed through social media. A subsequent survey of 56 Igorot migrants revealed that popular social media has a significant role in the exchange, revitalization, practice, and learning of indigenous culture; inciting an effective medium to leverage preservation strategies.*

## 1. Introduction

The preservation of indigenous knowledge has been recognized as a vital part of the sustainability of indigenous human societies in this age of globalization [30]. However, participation of indigenous peoples in migration and their exposure to the mainstream elements of modernization have brought challenges [35]. First, indigenous culture[1] is overwhelmed through assimilation into a different dominant mainstream culture [41]. Second, mass media culture could gradually replace peoples' indigeneity considering exposure to predominant media and social media [25]. Third, people are mesmerized by modern technologies and this buries their traditional relationship with the natural world [11]. The concern of losing indigenous knowledge revolutionizes efforts towards preservation.

The universal goal is its retention among community members and its transmission from the present generation to the next [16]. Accordingly, there have been a considerable number of significant ICT-based approaches implemented to help address this cause. They range from simple databases to massive digitization projects such as e-libraries, e-museums, comprehensive websites, information systems, and knowledge systems; several examples are enumerated and discussed in [39], [38], and [23]. Yet, these novel works could interface with the popular social media to further preservation outcomes.

In recent years, Information and Communication Technologies (ICT) have revolutionized the proliferation of social media platforms where individuals and communities create, share, and discuss content [22]. Social media have become a phenomenon where users can continuously modify contents and applications on the web in a participatory and collaborative way [20]. This phenomenon has enabled the creation of numerous online applications that support user generated content. Examples of such applications are blogs, collaborative projects such as Wikipedia, social network sites such as Facebook, and content communities such as YouTube. Thus, the emergence of social media has become a new opportunity to realize the aims of preservation [2].

This study explores the role of Facebook groups in sustaining indigenous knowledge among Igorot migrants. Despite being out of their traditional homes and living in a different society, they have maintained community cohesiveness and unyielding connection to traditional culture, which they manifest through the groups. This observation confutes the notion of the likely loss of culture due to their involvement in the activities of modernity.

---

[1] The terms *culture, indigenous knowledge, customs, heritage,* and *traditions* refer to the same context in this study and are used interchangeably throughout the text.

There have been studies conducted on other indigenous groups regarding the use of social media to negotiate indigeneity. The studies of Harris and Harris [16] and SanNicolas-Rocca and Parrish [41] compared the use of social media against traditional (e.g. books, oral) and ICT-based (e.g. radio, television) medium of transmissions of indigenous knowledge. Both have indicated the popularity of use of social media among indigenous peoples; however, they did not detail the extent to how social media facilitates preservation of knowledge. On the other hand, the ethnographic approach of Longboan [26] studied the longest running online forum of Igorots –*bibaknets@yahoogroups.com*, that has been using online narratives to assert the collective indigenous identity among members. However, members of this forum are dominated by the over-50s and their discussions have been significantly dwindling [26]. Likewise, online forums have been overwhelmed by newer social networking sites, which accommodate multi-media and visual content. Hence, there is a need to explore new and popular social media entities to discover what opportunities they offer for preserving indigenous knowledge.

## 2. Study background

### 2.1. Indigenous knowledge

Indigenous knowledge (IK) is deeply-rooted to generations of evolving practices of indigenous peoples [1], generally encompassing intellectual and cultural creativities that have defined their abilities and well-being. The broad scope of indigenous knowledge, also referred to as traditional or local knowledge, includes languages; cultural heritage in the forms of traditional stories, songs, dances, and ceremonies; and rituals that reflect beliefs related to spirituality, family, land, and social justice [41], [17]. It also includes, but is not limited to dwellings, arts, traditional sacred sites, oral history, food, traditional medicine, and clothing. Roy [38] has even classified IK into two, based on physicality: tangible and intangible representations. Indigenous knowledge, which have been internally accumulated and developed through time, are distinct from those of external mainstream society or other sections of the national or global community.

The value and relevance of indigenous knowledge has been recognized as an important factor for the survival of human societies [30]. This knowledge reinstates people's closer relationship to the natural world; a relation that went through millions of years of evolutionary development but is being suppressed by the elements of modernity[11]. Likewise, it is considered the soil that provides a society's nourishment and the basis to define value systems, behaviour, morals, ethics, and its peoples' future [36]. On this basis, there have been significant global movements among concerned organizations and communities to preserve endangered indigenous knowledge for future generations [39], [23]. In these undertakings, ICT has been one of the dominant tools in the capture, preservation, and transportation of indigenous knowledge.

### 2.2. The Igorots as indigenous peoples

Generally, Indigenous Peoples (IP) are considered aboriginal or native to the lands they live in. As stated in the Indigenous and Tribal Peoples Convention of 1989, people to be considered indigenous are either: *i)* the descendants of those who inhabited a geographical area before colonization, or *ii)* they have maintained their own social, economic, cultural and political institutions since colonization and the establishment of new states [18]. The IWGIA (International Work Group for Indigenous Affairs) reported that 370 million people worldwide are considered indigenous, mostly living in remote areas of the world [19].

The Igorots, also referred to as *highlanders* [12], denoting "people from the mountains" [29], is a term that collectively refers to the ethnolinguistic groups (*tribes*) of the mountainous Cordillera Region (CAR) located in the northern part of Luzon in the Philippines. This term has particularly been used by most local and international historians and researchers in their works [28] [12] [10] [21] [42]. The Igorots are historically differentiated from the majority of Filipinos because of their strong and successful resistance to colonization [26]. Their isolation and autonomy from centuries of Spanish colonial rule was a catalyst to sustain their indigenous customs, livelihoods, and access to communal lands throughout time [4]. Thus, they become deeply rooted to their culture and they have extensively continued their practices regardless of the new influences of modern societies.

### 2.3. Migrant Igorots and challenges to indigeneity

The United Nations Global Migration statistics report revealed that in 2013 there were 232 million international migrants [44]. In another report published by the Commission on Filipino Overseas, there were an estimated 10.49 million Overseas Filipinos (OF) in 2013 distributed in about 218 countries and territories around the world [6]. The latter report explains that the OFs comprise three groups: the *permanent migrants*, whose stay does not depend on employment and are legitimate permanent residents overseas; the *temporary migrants*, whose stay overseas is for employment or study-related and are expected to return to the Philippines after their contract; and the *irregular migrants*, whose stay overseas is not properly documented. It also cited that permanent migrants (47%) comprised the largest category of OFs, followed by temporary migrants (40%) and irregular migrants (13%). The same CFO report shows a total of nearly 17,500 permanent migrants whose place of origin is the Cordillera Region. Likewise, in a report made by the Philippine Statistics Authority, there were more than 50,000 temporary migrants from the Cordillera Region in 2014 [34]. Generally, these statistical reports from the different organizations depict how migration is becoming widespread around the world, and this phenomenon does not exclude the indigenous groups; moreover, the reports have shown a significant number of Igorot migrants.

Previously isolated from modernity, the indigenous groups have become participants to various processes of globalization [3]. Berry [3] argues that the complex process of globalization starts when societies engage in international contact, a process that involves a flow of cultural elements and establishment of relationships and networks. Regardless of the numerous forms of resistance of some indigenous groups to the activities of globalization [15], it has been observed that large numbers of indigenous peoples have been exploring and connecting to the global society through migration. For instance, the globalization of the market economy, which has created a high demand for workers to provide labor in more-developed nations [32], has brought opportunities for Igorots to go out of their homes; they have seen migration as a means to uplift their economic well-being. Besides, some Igorots perceive it as an escape route to a place where they would no longer be considered underdeveloped and backward [27]. Therefore in this case, they become more proximate to various influences to the dynamic behaviour of culture [41]. These influences elicit the evolution of culture, which through time, may lead to the loss of some significant cultural traits. Nevertheless, Igorot migrants have shown that they have maintained strong ties to their heritage; their self-knowledge, identity, and community practices are even intensified outside their home village [27]. For instance, cultural events that showcase Igorot traditionalism are celebrated in various communities overseas [26].

### 2.4. The role of ICT and social media

Information and Communication Technologies (ICT) have laid the ground for the proliferation of worldwide communication and digitization, and indigenous peoples have taken advantage of its new products and technologies. The communication technologies provided by the Internet including their visual reliance to become effective, is not very different from traditional forms of indigenous communication. It follows that the Internet, being grounded to the traditional oral and visual forms of communication where the sense of community is immediate, could be rapidly adopted by most indigenous peoples [43]. According to Landzelius [23], "some indigenous groups are putting state-of-the art communications to use in the service of traditional practices and cultural forms" (p. 9). These endeavours of adopting ICT towards preservation of IK are detailed in the works of Lieberman [25], Sahoo and Mohanty [39], Roy [38], and Landzelius [23]. It is observed among the mentioned works the complete participation of indigenous communities throughout the process; a concrete approach in deliberately considering the careful blending of ICT with indigenous knowledge [25].

Social media applications, in particular, employ the pervasive mobile and web-based technologies to become highly interactive platforms via which individuals and communities share, create, discuss, and modify user-generated content [22]. These applications are classified into six groups by Kaplan and Haenlein [20] based on the levels of media richness of the content and self-presentation of the user: collaborative projects, blogs or forums, content communities, social networking sites, virtual game worlds, and virtual social worlds. The sense-of-community that most social media applications possess elicited their popularity of use. Hence, initiatives to use social media platforms as means of promoting indigenous culture are continuously growing. Live And Tell is a community-based content management system leveraged from social media technologies to create an environment for sharing, teaching, and learning the endangered language and culture of the Lakota peoples of North America [2]. The work of Greyling and McNulty [14] describes an online indigenous digital library as part of the public library services that involved active participation of the indigenous community to contribute content. Its wiki platform and the use of folksonomies allowed flexibility on categorizing contents. Meanwhile, the *bibaknets* on yahoogroups.com, the longest running online forum among Igorots, has been using online narratives to assert the collective indigenous identity among members [26].

Facebook, being one of the popular social media applications today [7], offers features for building virtual communities among dispersed indigenous peoples. It has become attractive to present one's own unique culture through Facebook, especially for the opportunities it offers for visual expression [46]. There have been studies conducted on the prevalent use of Facebook among indigenous groups. Harris and Harris [16] compares traditional physical influences against the influences of three ICT media – radio, TV, and Internet, in the transmission of traditional cultures among young Kelabit indigenous peoples. The study of SanNicolas-Rocca and Parrish [41] shows the extent of sharing and preserving indigenous knowledge through different ICT platforms (i.e. including social media) among the Chamorro peoples. Their study reveals that Facebook belongs to the top social networking websites that are used by Chamorros to learn and share cultural knowledge. However, they did not show the extent social media facilitates the learning. The ethnography of Virtanen [46] discusses how Facebook has become a major domain of communication and expression among Brazilian Amazonian indigenous groups. She further concludes that social media is not only a means of self-determination, but also generates new reflection of indigenous knowledge.

Based on the revealed popularity of social media among indigenous peoples in their communication and self-expression, we aim to examine its potential role towards preservation of knowledge. We extend the previous works by exploring indigenous peoples' social media engagements to determine effects to both revitalization of indigeneity and accumulation of knowledge. In particular, we focused on the Igorot migrants' use of Facebook groups and the groups' probable influence to sustaining their indigeneity and learning their culture.

## 3. Research method

Influenced by Yin [47] and Noor [31], we followed an exploratory case-study methodology using a post-positivist approach. This approach is relevant for the following reasons: *i*) the study involves an empirical investigation of an emerging phenomenon, which is the use of social media in negotiating indigeneity among the migrant Igorot peoples; *ii*) the study explores the hypothesis ─ that social media revitalizes indigeneity among the migrant Igorot peoples; *iii*) the study gathers facts to objectively measure occurrence of certain patterns; and *iv*) the study appreciates different interpretations of people on their experiences. In addition, it is important in the gathering and interpretation of data to involve an interpretivist approach [13]. Hence, affiliation and familiarity with the Igorot community has been essential to the study (i.e. one of the authors is an Igorot who has adequate orientation to the culture and whose membership to the Facebook groups became necessary).

We subdivided the overall research question – How do Igorot Facebook groups facilitate the preservation of indigenous culture among Igorot migrants? – into three sub-questions: *i*) What cultural elements are conveyed in Igorot Facebook groups? *ii*) Do the Facebook groups help revitalize one's indigeneity? and *iii*) Do the Facebook groups contribute to one's learning of culture? Then to address sub-question (*i*), we selected 20 Igorot Facebook groups (see Table 1, data was gathered on April 1-3, 2016) and assessed each group on the different cultural elements conveyed based on a rigid evaluation of wall posts (e.g. in the form of text, photos, videos, links, and statuses). The evaluation covers wall posts made from January 1, 2016 – March 31, 2016. Within the defined three-month period, aided by the Facebook group monitoring tool *sociograph.io*[2], we counted the occurrences of these elements in each group's wall posts. Subsequently, to address sub-questions (*ii*) and (*iii*), we conducted an online questionnaire survey. The link to the survey was posted on the groups' walls and was sent as a private message to randomly picked group members identified as migrants.

## 4. Result and discussion

### 4.1. Survey of Facebook groups

Starting from an initial set of IK classifications derived from insights gained from [41], [38], and [17], it evolved into 18 classifications to which each cultural element observed from the groups were mapped correspondingly.

a) **Values** – inherent to the culture of Igorots is a value called *binnadang* [24], a traditional practice of helping especially those in need; typically done through volunteering yourself in a community work or contributing any resources you have in times of others' need. This was described by Leeftink [24] as a non-reciprocal help which comes from the heart; it is voluntary, immediate, direct, and automatic. The practice of this value has been observed in several posts from the groups like participation in community works, money contributions for a cause, and announcements of attending a person's wake.

b) **Cultural memes** – these are personalised pieces of media, usually a graphic with a catchphrase or sometimes in a form of mimicry, which are posted online to depict a cultural idea, concept, or activity. Most of these memes establish fondness to the Igorot culture.

c) **Textual compositions** – these are in the form of essays, poems, and rhetorically or artistically composed statements to promote or defend one's indigeneity. Most of the posted compositions talk about stereotyping

---
[2] www.sociograph.io

Table 1. Surveyed Igorot Facebook groups.

| ID | Group name | *No. of members | No. of admins |
|---|---|---|---|
| 1 | Igorots/Cordillerans got talent/ Talentadong Igorot/Cordilleran | 69, 400 | 24 |
| 2 | I LOVE Cordillera Group | 56, 100 | 27 |
| 3 | Rambak Cordillera | 37, 500 | 14 |
| 4 | Igorots @ Facebook | 32, 800 | 7 |
| 5 | Benguet in Pictures | 25, 000 | 8 |
| 6 | Igorot Ak | 22, 200 | 3 |
| 7 | Cordillera (beautiful as paradise) | 21, 850 | 9 |
| 8 | Igorot Clan (Tawid a Kultura, Tradisyon ken Kannawidan) | 21, 000 | 9 |
| 9 | Sons and Daughters of Cordillera | 14, 800 | 4 |
| 10 | Proud of Igorots (Shy kito umali kayo sina tako) | 6, 800 | 10 |
| 11 | InterActiveCordillera | 4, 100 | 8 |
| 12 | Cordillera ay Kalalayadan | 3, 950 | 12 |
| 13 | ICS The Igorot Country Sound | 2, 780 | 9 |
| 14 | I'm proud to be Igorot (Kailyan esna tako man-istorya) | 1, 800 | 4 |
| 15 | Bibaknets | 1, 480 | 1 |
| 16 | Full blooded Igorot by heart | 1, 200 | 1 |
| 17 | Igorot kami | 1, 100 | 1 |
| 18 | Igorotak | 1, 100 | 2 |
| 19 | Dap-ayan di Cordillera South of France (DCSOF) | 220 | 6 |
| 20 | Igorot Norway | 80 | 10 |

*Rounded off to the nearest tens

experiences, sentiments of being an Igorot and strong statements of emphasis on being proud of being one.

d) **Arts** – this includes posts of ancestral dwellings, indigenous tools, weapons, sculpture, woodcarvings, body tattooing, and paintings. Igorots are known for various forms of artworks: the Ifugaos, which was described by Roxas-Lim [37] as people actively engaged in arts, particularly carving, weaving, and blacksmithing; the Kalingas, known for body tattooing [40]; and the art of mummifications practiced by the indigenous Benguet peoples [40].

e) **Language** – the use of native dialects (e.g., Ilokano, Kankana-ey, Ibaloi) in posts, conversations, and comments is prevalent in the groups. Although Ilokano may not be a native dialect of Igorots, it is used between tribes of different speaking tongues. Thus, a majority is observed using Ilokano in their posts and comments. However, it is also noticeable that the use of the English language is high. This attests to what McKay [28] mentioned that one source of pride of the Igorots is their proficiency in the English language. Furthermore, the use of indigenous language is seen in local songs and movies.

f) **Sceneries and places** – one of the most frequent posts and most-liked posts among the groups belong to this category. These posts depict the uncountable sceneries in the Cordillera Region including the popular Banaue Rice Terraces, Mt. Pulag, Sagada Caves, Baguio City, the numerous scenic mountains and falls of Bakun, and the panoramic places of Kalinga.

g) **Songs, music, and music videos** – it is clearly manifested from the survey that Igorots love their songs and music. Shares and posts of traditional and newly composed Igorot songs fairly dominate some groups' walls. Most posted items in this category are locally produced music videos, personal covers of songs, and live shots from live performances. It is noticeable that most contemporary Igorot songs and music videos still depict traditional settings and practices.

h) **Dances** – the performance of indigenous dances is one of the most renowned ways to highlight the Igorot identity. It displays a wide range of cultural elements including traditional practices, rituals, costumes, indigenous arts, and music. Although, each Igorot tribe has its own-called dances, a generically-called *tayaw* performed by people in a circular motion in harmony with the beat of gongs, has become the signature dance of Cordilleran identity [33]. Numerous posts in this category captured overseas Igorot communities performing traditional dances.

i) **Traditional ceremonies and rituals** – this category includes the variety of traditional practices and rituals uniquely associated to the beliefs of each tribe in the Cordilleras. Generally, these practices are reflected in major stages of an individual's life cycle (birth to death) and in the annual round of community activities [9]. Animal sacrifices, like chicken, pigs, and cattle, are central to Igorot rituals [8]; usually depending on the type of ritual (individual or community) and perhaps the "direction" of the so-called elders. We argue that despite the obsolescence of most traditional rituals, modern community-involved ceremonies still insinuate some facets of traditional practices (e.g. *cañao*). Thus, we included in this category posts regarding any community gatherings, implying even the partial enactment of traditional ceremonies like in weddings and family reunions.

j) **Traditional stories, narrations, and local movies** – this category includes stories of known Igorot mythology characters or deities like *Lumawig*, *Bugan*, *Kabigat*, and *Kabunian*; popular tales and legends; origin myths; cautionary tales; and nursery tales. Likewise, this category includes the satirical stories of humor among different tribes (e.g. *iKiangan* stories) and satire of Igorot experiences on encounters with other societies. We also included in this category posts about local movies.

k) **Traditional attire, clothing, and garments** – this includes posts exhibiting traditional clothing. Among most migrants, their boldness in wearing traditional clothing is apparent in most posts, particularly during celebrations and special events organized within their communities. Posts showing performances of traditional dances and participation in Igorot events entail wearing of traditional attire.

l) **History** – this comprises historical articles pertaining to Igorot peoples and the Cordillera Region. Captioned historical photos and articles about historical origins of cultural elements are also included. However, several photos showing historic events, places, and people are not labelled, which probably makes it difficult for members to appreciate.

m) **Food** – compared to other Filipino societies, there are only a few known traditional foods associated with the Igorots, which include *pinikpikan*, *etag*, *watwat*, and *tapey*. We also included in this category some posts on local indigenous produce or traditional fruits like *kamote* and *tugee*.

n) **Traditional medicine** – there are very rare occurrences of posts among all surveyed groups regarding traditional medicine and its practice. Some of the few posts noted from the groups are about the benefits of walking, medical cures of guava leaves and descriptions of some medicinal plants found in the Cordillera region.

o) **Environmental practices and concerns** – posts about concerns on rivers (e.g. Chico River), planting trees, and mining are found in the groups.

p) **Traditional livelihood practices and concerns** – posts regarding livelihood activities, traditional mining, sustainable farming, and organic farming practices are mapped to this category.

q) **Festivals and traditional sports** – cultural themed festivals are frequently held almost everywhere in the Cordillera Region and among Igorot communities abroad. Community festivals exhibit most facets of culture such as performances of traditional dances, sports, and traditional community practices. Facebook groups facilitate the space where captures of these events are posted and shared among communities. Examples of festivals posted in the groups are the Etag Festival of Sagada, Panagbenga Festival of Baguio, Lang-ay Festival of Bontoc, Adivay Festival of Benguet, Hagiyo Festival of Banaue, Lubuagan-Tabuk Festival, Am-among Festival of Bontoc, Ullalim Festival of Kalinga, and Gagayam Festival of Sabangan. Posts exhibiting traditional sports are also counted in this category such as *sanggol* (arm wrestling), *tursi* (finger wrestling) and *bultong* (belt wrestling of the Ifugaos).

r) **Current events** – this category comprises news and remarkable achievements of Igorot peoples in the local, national, and international scenes. Posts in this category are the highly rated. It gathered the largest volume of interactions (shares, comments, and likes) in the groups. Included in this category are posts about Igorot individuals who accomplished exemplary successes in academic studies, arts, education, sports, government, beauty pageants, show business, and in other fields. Based on the posts' contexts, Igorots regard such people as a source of pride.

*Discussion.* For each Facebook group, the level of occurrence of each IK classification was described based on $n$ – the number of cultural elements mapped to the classification. The description consisted of three levels: *high*, if ($n \geq 10$); *moderate*, if ($5 \leq n < 10$); and *low*, if ($n < 5$). We summarized in Figure 1 the levels of occurrence of each classification among the surveyed groups. For example, on sceneries and places, 50%, 20%, and 30% among the groups have respectively *high*, *moderate*, and *low* posts in this category. The extensive classification of IK and the observed frequency of posts prove that Igorot Facebook groups actively convey a comprehensive range of cultural elements. Conveying IK propels cultural transmission and we agree with Harris and Harris [16] that this transmission could initiate knowledge preservation.

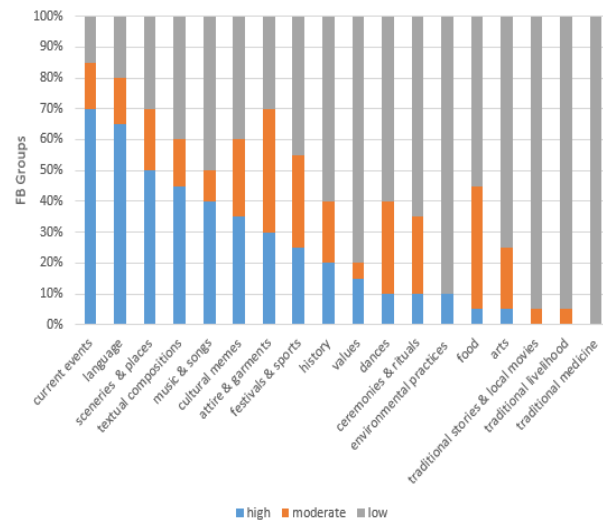

Figure 1. Level of occurrence of conveyed indigenous knowledge.

## 4.2. Survey of Igorot migrants: demographics

There were 56 respondents (42 females and 14 males) in the online questionnaire survey. All respondents claimed to be Igorot migrants and members of an Igorot Facebook group. Fifty-one (91%) considered themselves full-Igorots – both parents are Igorots, four (7%) are half-Igorots – either parents are Igorots, and one respondent (2%) stated having an Igorot lineage. In terms of migration status, 52% are temporary migrants and 48% are permanent. Based on their responses, the distribution of their locations is shown in Figure 2. Most respondents are migrants in New Zealand and Canada. Likewise, most respondents (42.9%) are overseas for only 0 to 3 years, while the rest have been overseas from four to more than 21 years. Moreover, the respondents also represented a wide range of ages from 20 below to 60 above, although 73.2% are within 21-40 years old.

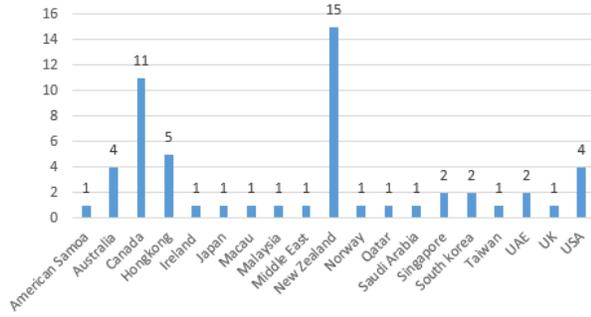

Figure 2. Location of respondents.

### 4.3. Engagement in Facebook groups

The levels of engagement of the respondents in Igorot Facebook groups are collectively shown in Figure 3. Four questions were asked to the respondents: *i*) about their frequency of visits to groups, *ii*) their levels of interaction (i.e. through likes and comments on group posts), *iii*) their likelihood of using native language in group interactions, and *iv*) their frequency of posting in groups (i.e. this includes sharing a post to groups). Generally, merging the four modes of engagement described shows that a majority of the respondents are frequently engaged in Igorot Facebook groups.

**Discussion.** Noticeable in Figure 3 are the 30% respondents indicating that they rarely post in the groups. This does not dispute the levels of frequency of posts shown in Figure 1, because posts can be from non-migrant members. Besides, this could agree with our observation that most posts are from group admins. In some groups, their liveliness depends on the engagement level of group admins. Likewise, most of the surveyed groups have several admins as shown in Table 1. We believe that admins play important roles in initiating group interactions and activities.

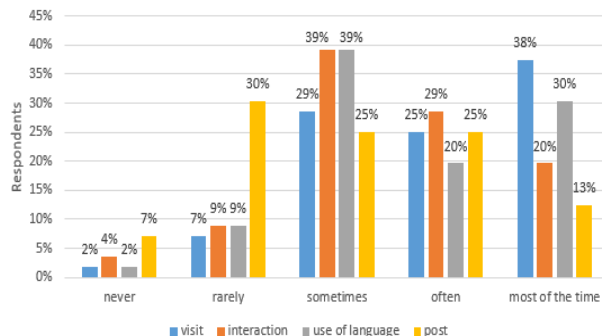

Figure 3. Frequency of visits, interactions, posts, and use of language.

### 4.4. Towards revitalizing indigeneity and learning indigenous culture

Firstly, respondents were asked to rate the influence of posting, sharing, reading, or viewing posts on Igorot Facebook groups in sustaining or uplifting their personal attachment to their Igorot culture. Secondly, they were also asked to rate how reading and viewing group posts contributed to their interest and knowledge of their Igorot culture. For answering both questions, the respondents used a five-point Likert scale: *very little*, *little*, *fair*, *high*, and *very high*. 45%, 32%, and 18% of the respondents indicated respectively a *fair*, *high*, and *very high* influence of Igorot Facebook groups to personal revitalization of indigeneity. Almost similarly, 39%, 32%, and 25% of the respondents said that Igorot Facebook groups contributed to their learning in a scale level of *fair*, *high*, and *very high* respectively.

In addition, the respondents were asked to choose (one) from a list of cultural elements, the most influencing post to their personal sense of indigeneity. The list consists of the top rated (i.e. has the most number of likes and comments) IK classification based on our observation from the preliminary survey (Section 4.1.). Most respondents (39%) considered news and current events the most influencing; dances, music, and sceneries got almost an equal share of respondents (13%-14%); followed by articles on history and culture (9%), then festivals (5%), and attire (2%). Furthermore, two respondents (4%) picked the option "other" (i.e. allows writing an answer not found in the options) and wrote "all".

**Discussion.** Based on these results, Igorot Facebook groups can play a significant role to one's indigeneity in two-folds: *i*) in terms of influencing revitalization (i.e. implied in this study as sustaining or uplifting attachment to indigenous culture) and *ii*) in terms of contributing to one's knowledge of culture. On the one hand, 95% of the respondents implied that Igorot Facebook groups have at least a *fair* influence to revitalizing their indigeneity. Revitalization is a process of sustaining or uplifting the sense of self (i.e. being Igorot). This process extends what Landzelius [23] called the growing language of diasporic indigenous migrants: the feeling to maintain, revive, and invent a connection with home. Thus in the case of the Igorot migrants, we argue that the Facebook groups can essentially contribute to this process. Moreover, the results support the theory of McKay [27] on the existence of a communications-technology-facilitated virtual village that enables migrants to connect, share, and express mutual feelings of their home village. This makes them remain at home in a global world, intensifying village social networks and expanding their senses of village-based self. We agree with McKay [27] that people do not practically leave their home village when they migrate; instead, they become much more engaged with the local identities and interpersonal of familial ties that make their village. Hence, the Facebook groups can facilitate the building of this intense connection and sustained sense of self, which could positively influence the preservation of indigenous knowledge. On the other hand, 96% of the respondents implied that Igorot Facebook groups have at least a *fair* level of contribution to their knowledge of culture. Knowledge about culture, awareness, appreciation, learning it, owning it, and practicing it are key values and processes for its preservation. We agree with Bisin and Verdier [5] that knowledge accumulation happens through experience, participation, and observation among members of the community. The findings in Section 4.3. indicate the potential of Facebook groups towards realizing these processes.

Moreover, the survey in Section 4.1 shows that Facebook groups offer interesting resources for learning the Igorot culture. Besides posts originating directly from members, there were valuable shared links to information websites. Several websites[3] cater collaborative information about Igorots and their culture, and Facebook groups became pro-active hubs to share and access those information.

### 4.5. Processes of preservation

From analysing the collected data and the results, and from the literatures used in this study, we identified candidate processes that the Facebook groups encompass for enabling preservation of indigenous knowledge. The identified processes evolved during the contextual analysis of the open-ended questions' responses. The responses were grouped into themes implied by their context. Consequently, the derived themes became the bases of the five processes shown in Table 2. We elicited different interpretations of the respondents on their experiences in the Facebook groups. Furthermore, these respondents' views became the bases of identifying the associated values developed through each process. We believe that values developed through the processes could become catalysts towards preservation of the indigenous knowledge.

There are values inherent to IK, as Ronchi [36] presumes that IK defines value systems and behaviour of peoples. However, we argue that in addition to the inherent values, there are also contemporary values generated from the ongoing social relationships and preservation activities. A notable example from the responses is the frequent mention of "pride". In the past, it would be awkward for someone to admit to being an Igorot [12], [29], [26], [10]. However, the exemplary deeds, achievements, practices, and collective expressions of Igorots circulated through social media, have gradually transformed a timid group into self-esteemed peoples. This illustrates what Valientes [45] described as ironic transformation: that by some twists of history, the "Igorot" has transformed to become a proud identity despite all the negative connotations that cling to the word.

As shown in Table 2, a value can appear in multiple processes based from our own contextualization of the perspectives. Likewise, the values are just representatives, can be expanded and will not be limited to the list. The table also shows only representative views related to each process. Moreover, as highlighted by the respondents, we do not discount the importance of physical gatherings, memberships to Igorot organizations, and physical involvements in the practice of Igorot customs, as primary means to preserve indigeneity. This is a notable quote from one respondent: *"Aside from social media, I can learn and preserve my Igorot root(s) by joining Igorot Organizations. Here in Vancouver, BIBAK is the most popular organization that they established to unite Cordilleran people. I was never active in joining cultural events since I was a child but when I came here, I learned different cultural danced(dances) and be(came) more proud of who I am."*

Community activities of Igorot migrants were evident in the documentations posted and shared in Facebook groups. Thus, social media became a tool to show-off exemplary practices that others would appreciate and replicate which could likewise contribute to sustainability and propagation of indigenous knowledge leading to its preservation.

## 5. Conclusion and future work

The interactive platforms and the sense-of-community property of social media applications, which allow users to share, create, discuss, and modify content, promote an environment of participation and collaboration. Studies on preserving indigenous knowledge advocate the complete participation of indigenous peoples in the process. We have revealed how the Igorot Facebook groups become resources of vast indigenous knowledge collaborated by members. At the same time, considering the vulnerability of Igorot migrants to the challenges of sustaining their indigeneity, the study reveals the potential complementary role of Facebook groups in the revitalization of indigeneity and learning of culture.

Facebook groups therefore can become alternative and complementing grounds to sustain indigenous knowledge. In the case of the Igorot migrants, the groups continue to provide a contemporary medium to

---

[3] a) http://www.cordilleransun.com/ b) http://igorotage.com/

Table 2. Different views of Igorot migrants on how Facebook groups preserve indigenous knowledge.

| Processes/Activities (values developed) | Perspectives |
|---|---|
| **Connection to culture** (friendship, unity, happiness, pride, knowledge, belongingness…) | *"To show how proud I am as an Igorot no matter how long I have been away from my birth country. To keep myself connected to my roots and build new friendships as well as learn new things about my tribe."* |
| **Facilitates learning** (awareness, expression, friendship, understanding, respect, knowledge…) | **"A good place of getting together. These growing kids are learning our culture through social media (shows its advantage). It is also where they can see the exemplary deeds of our fellow (Igorots) around the world."* |
| **Revitalization of indigeneity** (pride, confidence, appreciation, knowledge…) | *"Joining Igorot facebook group added my knowledge and information what is happening around the world to Igorot people. Sometimes I see people post how they celebrate Lang-ay Festival, their organization's anniversary etc. These helped me appreciate how Cordilleran people love and treasure their culture even if they are away from home."* |
| **Facilitates community participation and continuous practice of indigenous activities** (sharing, unity, belongingness, friendship, pride…) | **"The benefit of FB (group), is knowing what others (fellow Igorots) post about gatherings."* |
| **Information campaign** (pride, respect, knowledge…) | *"Facebook has been very helpful in showcasing the different cultures from different parts of the Cordillera not only locally but also worldwide. Through Facebook, many people were enlightened on what a Cordilleran or an Igorot looks like, the lifestyle and the like. Also, through these social media sites, people around the world are amazed on the unique talents and skills of a Cordilleran. And lastly, even fellow Cordillerans learn from each others culture with the use of social media."* |

*Translated*

establish virtual communities that connect them to traditional home, unite them with traditional culture, inform them about their fellows, and encourage them promote the knowledge. Furthermore, the groups could become pro-active learning resources for collaborated indigenous knowledge – a knowledge covering various elements of indigeneity, a knowledge represented in different media forms, a knowledge contributed and interpreted in varying perspectives, and a knowledge to reference model values.

Social media indeed has created a dynamic and attractive platform to transport the diverse cultural elements through various media forms. Its phenomenal use to actively promote indigenous knowledge makes it a vehicle for revitalization [23]. Likewise, pervasive technologies such as the mobility of smart devices integrating multimedia and internet services has enabled the omnipresent captures of knowledge artefacts and cultural exhibitions that could be instantly propagated and shared through social media, not just among a concerned group, but globally. We see two tracks to further this study. First, the opportunities for preservation strategies offered by the popular, unstructured, and dynamic social media could be leveraged towards a social-media-based framework for a more effective preservation of indigenous knowledge; while considering the limitations of social media on the accuracy, reliability, and control of conveyed information. Second, an extensive evaluation of the impact of the use of social media in the context of IK preservation, considering both Igorot peoples in the homeland and in the diaspora; likewise, to the younger generations of Igorots who are raised overseas.